\documentclass[9pt,twocolumn,twoside]{pnas-new}

\templatetype{pnasbriefreport} 

\title{Seabird trajectories map onto a reduced optimal-control bound for dynamic soaring}

\author[a,b,c]{Louis Gonz\'alez}
\author[d]{John P. Y. Arnould}
\author[a,b,c,1]{Saad Bhamla}

\affil[a]{School of Chemical \& Biomolecular Engineering, Georgia Institute of Technology, Atlanta, USA }
\affil[b]{BioFrontiers Institute, University of Colorado, Boulder, USA}
\affil[c]{Department of Chemical \& Biological Engineering, University of Colorado, Boulder, USA}
\affil[d]{School of Life and Environmental Sciences, Deakin University, Victoria, Australia}

\leadauthor{Gonz\'alez}

\authorcontributions{Conceptualization: L.G., Methodology: L.G., Data Analysis: L.G., Investigation: L.G., Data Collection: J.A., Supervision: S.B. Writing—original draft: L.G., Writing—review and editing: L.G., S.B.}
\authordeclaration{The authors declare no competing interests.}

\correspondingauthor{\textsuperscript{1}To whom correspondence should be addressed. E-mail: saad.bhamla@colorado.edu}


\begin{abstract}
\vspace{-.5cm}
Dynamic soaring allows seabirds to harvest mechanical energy from vertical wind shear, yet there is no common  benchmark for comparing flight performance across species from their trajectories. We derive a reduced lower bound on transport effort from a simplified Hamilton-Jacobi-Bellman optimal-control model in which slow flight incurs an induced-drag penalty, fast flight incurs a dissipative penalty, and wind shear supplies an effective energetic subsidy. \add{We rescale  each of four species to its own baseline speed  and  accelerometer-based effort, then map them onto a common reduced speed--effort plane and estimate each one's lower frontier. We calibrate the optimal-control bound to one species, the wandering albatross, and test the other three against it. Two further dynamic soarers, the Buller's albatross and short-tailed shearwater, lie progressively above the bound. The common crane, a thermal soarer of comparable body mass, lies about 33 times farther from it than the albatross. Proximity to the bound therefore measures how much of a bird's transport is powered by wind
shear. More generally, our  work offers a framework for testing optimal-control limits in bird flight using field data.}

\end{abstract}

\doi{\url{www.pnas.org/cgi/doi/10.1073/pnas.XXXXXXXXXX}}

\newcommand{\add}[1]{#1}
\setboolean{displaywatermark}{false}

\begin{document}

\maketitle
\thispagestyle{firststyle}
\ifthenelse{\boolean{shortarticle}}{\ifthenelse{\boolean{singlecolumn}}{\abscontentformatted}{\abscontent}}{}
\firstpage[9]{4}

Dynamic soaring allows birds to extract mechanical energy from vertical wind shear and sustain long-range flight with little active  propulsion~\cite{taylor2016soaring,bousquet2017optimal}. This flight mode is especially prominent in Procellariiform  seabirds~\cite{richardson2018flight,bousquet2017optimal,yonehara2016flight,richardson2022observations}, where wandering albatrosses can cover more than 10,000 km in a single foraging trip.  Recent work has shown that birds exploit wind shear in structured and often optimized ways~\cite{kempton2022optimization,chen2025optimal}, but field data still lack a benchmark that places observed trajectories relative to a minimum-cost boundary.

\add{To assess such a boundary, we analyze GPS and accelerometry  from four bird groups that  range from dynamic soaring specialist to a thermal soarer. The wandering albatross (\textit{Diomedea exulans}), a Southern Ocean shear specialist, contributes 55 deployments from 44~individuals~\cite{uesaka2023wandering};  the Buller's albatross (\textit{Thalassarche bulleri}, 32 deployments)~\cite{poupart2019bullers}, and  short-tailed shearwater (\textit{Ardenna tenuirostris}, 13 deployments)~\cite{berlincourt2015combined} are procellariiforms that  mix flap-gliding with wind-shear soaring; and the common crane (\textit{Grus grus}, 32 individuals)~\cite{pekarsky2024cranes} is a thermal soarer of comparable mass to the wandering albatross, included as a cross-mechanism contrast.} From each dataset, we derive a ground-referenced transport speed and an accelerometry-based effort measure, normalize each species to its own baseline, and map them onto a common reduced speed--effort plane. 




%

\subsection*{Albatross trajectories define an empirical transport--effort frontier consistent with wind-assisted flight}

\begin{figure*}[ht]
    \centering
    \includegraphics[width=\linewidth]{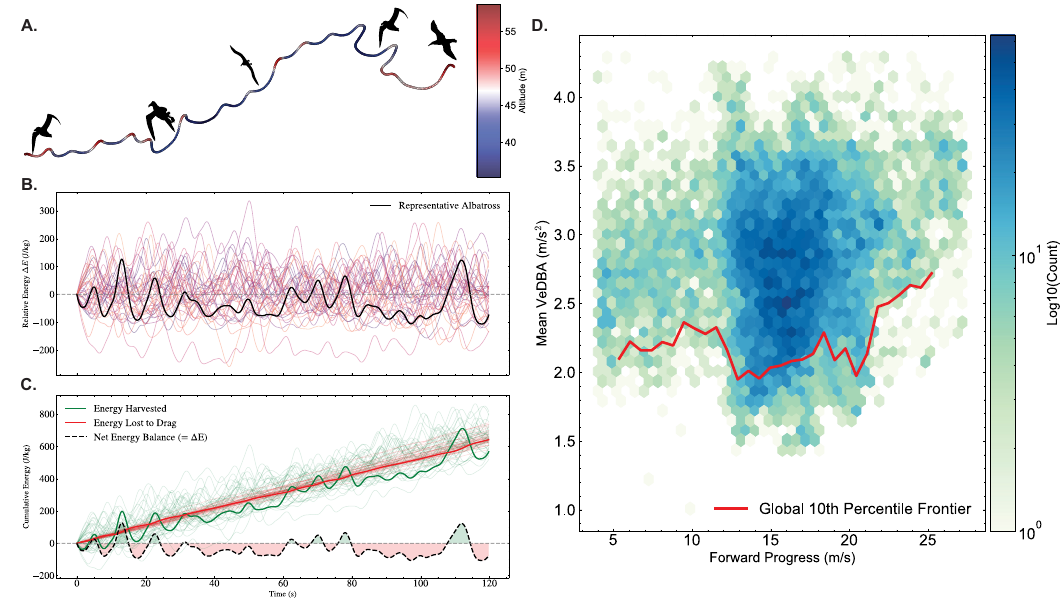}
    \caption{\textbf{Albatross trajectories show cyclic energy exchange and define an empirical transport--effort frontier.} \textbf{A.} Representative wandering albatross, colored by altitude, showing repeated turning and climb--descent cycles typical of dynamic soaring. \textbf{B.} Relative ground-frame specific mechanical energy $\delta E(t) = E(t) - E(t_0)$ for multiple 120-s trajectory windows, with one representative window highlighted. \textbf{C.} Cumulative energy harvested, $W_{\mathrm{harvest}}(t)$, and energy lost to drag, $W_{\mathrm{drag}}(t)$, whose difference is the net energy change $\Delta E = W_{\mathrm{harvest}} - W_{\mathrm{drag}}$ under the gliding approximation.
     \textbf{D.} Global hexbin density in the plane of net transport speed and mean VeDBA, with the red curve denoting the global 10th-percentile lower frontier estimated by quantile regression (SI\,S9).  }
    \label{fig:figure1} 
\end{figure*}

Fig.~\ref{fig:figure1}a displays a representative albatross trajectory, colored by altitude, with the repeated turning and climbing-descent cycles characteristic of dynamic soaring.  Continuous tracks were partitioned into 120-s overlapping windows \add{(step 30\,s; see SI\,S8 for window-length sensitivity across 60--240\,s, which yields an identical species ordering)}, and for each window, we computed the ground-frame specific mechanical energy $E(t)=\frac{1}{2}\,u(t)^{2}+g\,z(t)$, where $u(t)$ is ground speed and $z(t)$ is altitude. Figure~\ref{fig:figure1}b plots the relative energy $ \delta E(t)=E(t)-E(t_0)$ across many windows, with one representative trace highlighted. The oscillatory pattern reflects repeated exchanges between kinetic and potential energy during sheared flight.

To interpret these fluctuations, we estimate a cumulative drag work term using a quasi-steady gliding model and infer the corresponding cumulative atmospheric-input term (or wind-mediated gain)
$W_{\mathrm{harvest}}(t)=\delta E(t)+W_{\mathrm{drag}}(t)$, where $W_{\mathrm{drag}}(t)\ge 0$ is the cumulative drag loss (SI\,S2), under the approximation that sustained muscular input is small over these gliding-dominated windows. Figure~\ref{fig:figure1}c shows these two contributions, confirming them to be of comparable magnitude and opposite sign, consistent with a near-steady energy budget maintained by shear exploitation~\cite{taylor2016soaring,yonehara2016flight}. Because the ambient wind field is not reconstructed pointwise, $W_{\mathrm{harvest}}$ is interpreted as an effective atmospheric-input observable rather than a direct measurement of aerodynamic wind work. Its magnitude is nonetheless comparable to the inferred drag loss, consistent with atmospheric energy input offsetting dissipation during sustained transport.



\begin{figure*}[ht]
    \centering
    \includegraphics[width=\linewidth]{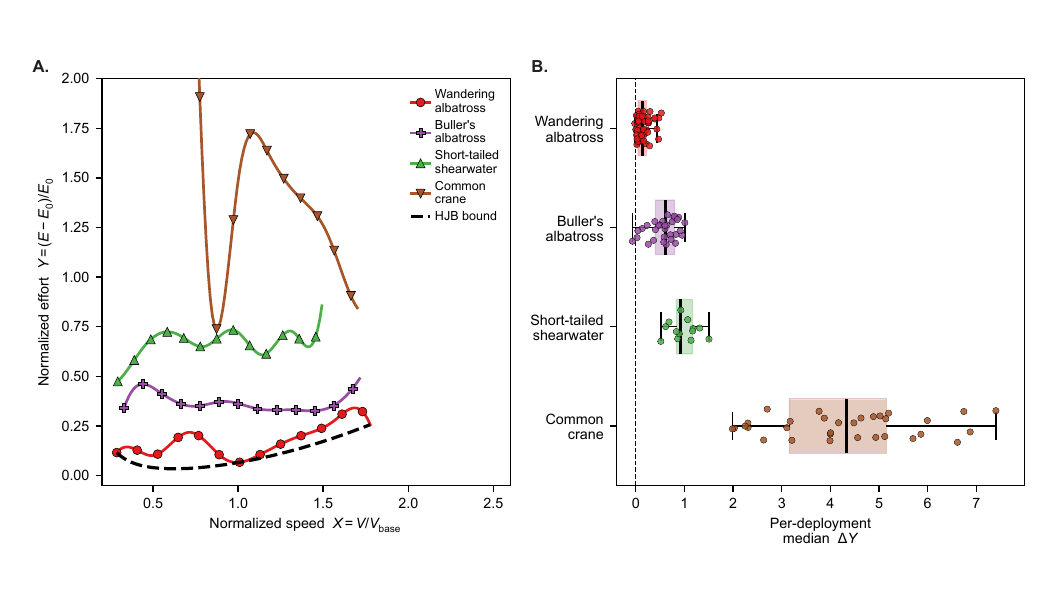}
    \caption{\add{\textbf{A reduced bound discriminates among soaring mechanisms.} \textbf{A.} Species-specific 25th-percentile lower frontiers in the reduced speed--effort plane. The dashed curve is the reduced HJB bound  fit to the wandering albatross frontier. The  procellariiform frontiers lie at progressively   higher residual effort, and the common crane, a thermal soarer of similar mass, lies well above them. \textbf{B.} Per-deployment median vertical offset  $\Delta Y = \mathrm{median}(Y_{\mathrm{obs}} - Y_{\mathrm{HJB}}(X))$, within the albatross-supported $X$-range. Each point is one deployment; boxes show the median and IQR. The wandering-albatross-to-crane separation ($\approx 33\times$) shows that distance from the bound reflects the mechanism of atmospheric energy extraction (wind shear vs.\ thermals) rather than soaring capability.} }
    \label{fig:figure2} 
\end{figure*}

Fig.~\ref{fig:figure1}d then collapses all retained albatross windows into the empirical plane of net transport speed versus mean vectorial dynamic body acceleration (VeDBA), an accelerometer-based proxy for locomotor effort~\cite{qasem2012tri}. The hexbin cloud shows the population distribution, and the red curve marks the 10th-percentile lower frontier, estimated by quantile regression on a cubic B-spline basis (the same estimator as Fig.~\ref{fig:figure2}, at $\tau=0.10$; SI\,S9). This lower frontier serves as our key empirical reference:  it summarizes the best-observed transport-effort trade-off in the albatross dataset. It provides the albatross baseline for reduced cross-species comparison in Fig.~\ref{fig:figure2}. 


\subsection*{A reduced HJB benchmark organizes cross-species transport--effort frontiers}

To compare species with different absolute flight speeds and effort scales, we transformed each dataset into the reduced variables $X=V_{\mathrm{obs}}/V_{\mathrm{base}}$ and
$Y=(\mathcal{E}_{\mathrm{obs}}-\mathcal{E}_0)/\mathcal{E}_0$, \add{where $\mathcal{E}$ is VeDBA for the seabirds and the running 4-second wingbeat rate for the crane, and $V_{\mathrm{base}}$ and $\mathcal{E}_0$ are per-species low-effort baselines. Because $Y$ is a within-species ratio, any species-specific DBA-to-power factor cancels, so $Y$ measures relative effort rather than metabolic rate~\cite{conners2024dynamic}. For each species we estimated a lower frontier, the 25th conditional quantile of $Y$ given $X$. We calibrated the bound to the wandering albatross alone and treated the other three species as held-out tests; cross-validation confirms the calibration generalizes across the albatross population (SI\,S5, S10). Altitude enters only Fig.~\ref{fig:figure1}b,c, which use the wandering albatross
from a single recorder; the cross-species comparison does not.} The resulting species frontiers are shown in Fig.~\ref{fig:figure2}a.

The dashed curve is the reduced Hamilton-Jacobi-Bellman (HJB)~\cite{bousquet2017optimal,chen2025optimal} form
\begin{equation*}
Y_{\mathrm{HJB}}(X)=\left[\frac{a}{X^2}+bX^2-WX\right]_+,
\end{equation*}
whose functional form follows from a simplified optimal-control model of along-track transport in prescribed shear. In this reduced description, the $a/X^2$ term captures the penalty for slow flight, the $bX^2$ term captures the high-speed dissipative cost, and the $-WX$ term represents an effective wind-energy subsidy supplied by shear exploitation. Because $Y$ is excess effort over a low-effort baseline, $[\,\cdot\,]_{+}$ keeps it non-negative. The functional form is derived from the model, whereas the coefficients $(a,b,W)$ are estimated from the wandering albatross frontier \add{by constrained least squares, with the  curve held at or below every frontier point (a true lower bound), subject to $a,b,W \ge 0$ and $W \ge 0.05$, a physical floor that keeps the wind-subsidy term from collapsing to zero and  contradicting the observed wind-energy extraction. Fit to all 55 deployments, $W$ settles marginally above this floor at $0.0577$}.

\add{Among the procellariiforms, the wandering albatross, a Southern Ocean shear specialist, sits closest to the bound, as expected for the calibration species, while the Buller's albatross and short-tailed shearwater, which flap-glide more but still exploit wind shear, sit at progressively higher residual effort
(Fig.~\ref{fig:figure2}a). We summarize each species by its per-deployment offset from the bound, $\Delta Y = \mathrm{median} (Y_{\mathrm{obs}} - Y_{\mathrm{HJB}}(X))$. Each deployment median averages over
many thousands of wingbeats, the scale at which VeDBA serves as a within-species relative-effort index. A small $\Delta Y$ marks flight near the minimum-cost regime, a large one a systematic departure. The three
dynamic soarers stay within about one unit of the bound ($\Delta Y \approx 0.13$, $0.61$, and $0.93$), while the thermal-soaring common crane sits far beyond at $\Delta Y \approx 4.33$, roughly $33\times$ the albatross (Fig.~\ref{fig:figure2}b). This ordering is unchanged when the window length is varied from 60 to 240~s (SI\,S8).}

\subsection*{Concluding Remarks} 

From Leonardo da Vinci’s account of birds in wind shear~\cite{richardson2019leonardo} to Rayleigh’s 1883 argument~\cite{rayleigh1883soaring}, to  modern analyses of soaring energetics, dynamic soaring has been treated as a problem in which atmospheric structure enters the transport budget of flight \cite{taylor2016soaring,bousquet2017optimal}. We reduce that problem at the scale of transport, so that  trajectories  expressed in normalized speed and effort place birds of different flight modes  in a common plane, where their  frontiers can be compared with the   HJB lower bound. \add{Distance from the bound measures how much of a bird's transport cost is recouped from wind shear. The common crane, a soaring bird of similar mass that rides thermals rather than shear, sits far above the procellariiforms, so the distance reflects the mechanism of energy capture, not soaring ability or body size. }


\add{The framework has three main limitations. First, because our energy balance is computed in the ground frame, where ground speed and airspeed cannot be separated, $W_{\mathrm{drag}}$ and $W_{\mathrm{harvest}}$ are effective measures rather than faithful aerodynamic decompositions. Second, the quasi-steady gliding model assumes a constant $L/D$ and ignores  posture, bank angle, Reynolds number, and transient loading. Third, the  HJB coefficients are fit to the albatross  frontier, so the curve  is an empirically calibrated benchmark rather than a prediction from aerodynamic theory.}
Our reduced  coordinates and optimal-control bound offer a  route from tracking archives to mechanistic inference, and  should extend  to  benchmarking engineered dynamic soarers  and reconstructing winds from bird trajectories~\cite{yonehara2016flight}.




  \vspace{\baselineskip}
\noindent \textbf{Data, Materials, and Software Availability.}  All study data are included in
the article and/or at an online data repository: \url{https://github.com/bhamla-lab/albatross-gonzalez-2026} and \url{https://zenodo.org/records/21304763}.

\vspace{-.3cm}
\acknow{S.B. acknowledges support from NSF awards PHY-2310691 and iOS-1941933; National Institutes of Health (NIH) award R35GM142588 and Schmidt Sciences, LLC.}

\showacknow{} 
\vspace{-.7cm}
\bibsplit[1]

\vspace{-1cm}
\bibliography{bibliography}

@article{richardson2018flight,
  title={Flight speed and performance of the wandering albatross with respect to wind},
  author={Richardson, Philip L and Wakefield, Ewan D and Phillips, Richard A},
  journal={Movement ecology},
  volume={6},
  number={1},
  pages={3},
  year={2018},
  publisher={Springer}
}

@article{conners2024dynamic,
  title={Dynamic soaring decouples dynamic body acceleration and energetics in albatrosses},
  author={Conners, Melinda G and Green, Jonathan A and Phillips, Richard A and Orben, Rachael A and Cui, Chen and Djuri{\'c}, Petar M and Heywood, Eleanor and Vyssotski, Alexei L and Thorne, Lesley H},
  journal={Journal of Experimental Biology},
  volume={227},
  number={18},
  pages={jeb247431},
  year={2024},
  publisher={The Company of Biologists Ltd}
}

@article{rayleigh1883soaring,
  title={The soaring of birds},
  author={Rayleigh, Lord},
  journal={Nature},
  volume={27},
  number={701},
  pages={534--535},
  year={1883}
}

@article{taylor2016soaring,
  title={Soaring energetics and glide performance in a moving atmosphere},
  author={Taylor, Graham K and Reynolds, Kate V and Thomas, Adrian LR},
  journal={Philosophical Transactions of the Royal Society B: Biological Sciences},
  volume={371},
  number={1704},
  year={2016},
  publisher={The Royal Society}
}

@article{bousquet2017optimal,
  title={Optimal dynamic soaring consists of successive shallow arcs},
  author={Bousquet, Gabriel D and Triantafyllou, Michael S and Slotine, Jean-Jacques E},
  journal={Journal of The Royal Society Interface},
  volume={14},
  number={135},
  year={2017},
  publisher={The Royal Society}
}

@article{yonehara2016flight,
  title={Flight paths of seabirds soaring over the ocean surface enable measurement of fine-scale wind speed and direction},
  author={Yonehara, Yoshinari and Goto, Yusuke and Yoda, Ken and Watanuki, Yutaka and Young, Lindsay C and Weimerskirch, Henri and Bost, Charles-Andr{\'e} and Sato, Katsufumi},
  journal={Proceedings of the National Academy of Sciences},
  volume={113},
  number={32},
  pages={9039--9044},
  year={2016},
  publisher={National Academy of Sciences}
}

@article{uesaka2023wandering,
  title={Wandering albatrosses exert high take-off effort only when both wind and waves are gentle},
  author={Uesaka, Leo and Goto, Yusuke and Naruoka, Masaru and Weimerskirch, Henri and Sato, Katsufumi and Sakamoto, Kentaro Q},
  journal={Elife},
  volume={12},
  pages={RP87016},
  year={2023},
  publisher={eLife Sciences Publications, Ltd}
}

@article{richardson2022observations,
  title={Observations and models of across-wind flight speed of the wandering albatross},
  author={Richardson, Philip L and Wakefield, Ewan D},
  journal={Royal Society Open Science},
  volume={9},
  number={11},
  year={2022},
  publisher={The Royal Society}
}

@article{kempton2022optimization,
  title={Optimization of dynamic soaring in a flap-gliding seabird affects its large-scale distribution at sea},
  author={Kempton, James A and Wynn, Joe and Bond, Sarah and Evry, James and Fayet, Annette L and Gillies, Natasha and Guilford, Tim and Kavelaars, Marwa and Juarez-Martinez, Ignacio and Padget, Oliver and others},
  journal={Science advances},
  volume={8},
  number={22},
  pages={eabo0200},
  year={2022},
  publisher={American Association for the Advancement of Science}
}

@article{chen2025optimal,
  title={Optimal dynamic soaring trades off energy harvest and directional flight},
  author={Chen, Lunbing and Yin, Yufei and Xiang, Yang and Qin, Suyang and Liu, Hong},
  journal={IScience},
  volume={28},
  number={6},
  year={2025},
  publisher={Elsevier}
}

@article{qasem2012tri,
  title={Tri-axial dynamic acceleration as a proxy for animal energy expenditure; should we be summing values or calculating the vector?},
  author={Qasem, Lama and Cardew, Antonia and Wilson, Alexis and Griffiths, Iwan and Halsey, Lewis G and Shepard, Emily LC and Gleiss, Adrian C and Wilson, Rory},
  journal={PloS one},
  volume={7},
  number={2},
  pages={e31187},
  year={2012},
  publisher={Public Library of Science San Francisco, USA}
}

@article{richardson2019leonardo,
  title={Leonardo da Vinci's discovery of the dynamic soaring by birds in wind shear},
  author={Richardson, Philip L},
  journal={Notes and Records: the Royal Society journal of the history of science},
  volume={73},
  number={3},
  pages={285--301},
  year={2019},
  publisher={The Royal Society}
}

@article{pekarsky2024cranes,
  title={Cranes soar on thermal updrafts behind cold fronts as they migrate across the sea},
  author={Pekarsky, Sasha and Shohami, David and Horvitz, Nir and Bowie, Rauri CK and Kamath, Pauline L and Markin, Yuri and Getz, Wayne M and Nathan, Ran},
  journal={Proceedings of the Royal Society B: Biological Sciences},
  volume={291},
  number={2015},
  pages={20231243},
  year={2024}
}

@article{berlincourt2015combined,
  title={Combined use of GPS and accelerometry reveals fine scale three-dimensional foraging behaviour in the short-tailed shearwater},
  author={Berlincourt, Maud and Angel, Lauren P and Arnould, John PY},
  journal={PloS one},
  volume={10},
  number={10},
  pages={e0139351},
  year={2015},
  publisher={Public Library of Science San Francisco, CA USA}
}

@article{poupart2019bullers,
  author  = {Poupart, Timoth\'{e}e A. and Waugh, Susan M. and
             Miskelly, Colin M. and Kato, Akiko and Angel, Lauren P. and
             Rogers, Karyne M. and Arnould, John P. Y.},
  title   = {Fine-scale foraging behaviour of southern {B}uller's albatross,
             the only \textit{Thalassarche} that provisions chicks through
             winter},
  journal = {Marine Ecology Progress Series},
  volume  = {625},
  pages   = {163--179},
  year    = {2019},
  doi     = {10.3354/meps13042}
}

@article{waugh2017muttonbirders,
  author  = {Waugh, Susan M. and Poupart, Timoth\'{e}e A. and
             Miskelly, Colin M. and Stahl, Jean-Claude and
             Arnould, John P. Y.},
  title   = {Human exploitation assisting a threatened species? The case of
             muttonbirders and {B}uller's albatross},
  journal = {PLoS ONE},
  volume  = {12},
  number  = {4},
  pages   = {e0175458},
  year    = {2017},
  doi     = {10.1371/journal.pone.0175458}
}

\end{document}


\title{Supplementary Information: Seabird trajectories map onto a reduced optimal-control bound for dynamic soaring}

\author[1,2,3]{Louis Gonz\'alez}
\author[4]{John P. Y. Arnould}
\author[1,2,3]{Saad Bhamla}

\affil[1]{School of Chemical \& Biomolecular Engineering, Georgia Institute of Technology, Atlanta, USA }
\affil[2]{BioFrontiers Institute, University of Colorado,
Boulder, USA}
\affil[3]{Department of Chemical \& Biological Engineering, University of Colorado, Boulder, USA}
\affil[4]{School of Life and Environmental Sciences, Deakin University, Victoria, Australia}
\vspace{-1em}

\date{}

\begingroup
\let\center\flushleft
\let\endcenter\endflushleft
\maketitle
\endgroup

\setcounter{tocdepth}{2}
\tableofcontents
\bigskip

\section*{Extended Methods}
\addcontentsline{toc}{section}{Extended Methods}

\sisection{S1. Mechanical framework for dynamic soaring}
We model the bird as an effective glider moving through an atmospheric flow. Let $\bm{x}(t)$ denote position in an Earth-fixed frame and let
\begin{equation}
\bm{u}(t) = \dot{\bm{x}}(t)
\end{equation}
be the ground velocity, and let $\bm{w}(\bm{x},t)$ be the local wind. The air-relative velocity is then
\begin{equation}
\bm{v}(t) = \bm{u}(t) - \bm{w}(\bm{x}(t),t).
\end{equation}
The specific mechanical energy in the ground frame is
\begin{equation}
E(t) = \frac{1}{2}\abs{\bm{u}(t)}^2 + g z(t),
\end{equation}
where $z(t)$ is altitude, and $g$ is gravitational acceleration. \add{The horizontal Doppler ground speed here is $\abs{\bm{u}}$. Vertical motion only enters through $gz$ and the kinetic term $\tfrac{1}{2}\dot{z}^{2}$ is thus dropped. For the wandering albatross that acts as our ideally efficient benchmark, the vertical speed has a median of $0.3$~m/s ($95$th percentile ${\sim}1$~m/s), keeping $\tfrac{1}{2}\dot{z}^{2}$ below $1\%$ of the horizontal kinetic energy in sustained flight.}

The translational equation of motion is written as
\begin{equation}
m\dot{\bm{u}} = \bm{F}_{\mathrm{aero}} + \bm{F}_{\mathrm{muscle}} - mg\,\hat{\bm{z}},
\end{equation}
where $\bm{F}_{\mathrm{aero}}$ is the total aerodynamic force and $\bm{F}_{\mathrm{muscle}}$ is the net propulsive force associated with active efforts. We note that ``active efforts'' here involve flapping or other types of muscular work. Taking the dot product with $\bm{u}$ gives
\begin{equation}
m\,\bm{u}\cdot\dot{\bm{u}}
=
\bm{u}\cdot\bm{F}_{\mathrm{aero}}
+
\bm{u}\cdot\bm{F}_{\mathrm{muscle}}
-
mg\,\dot{z}.
\end{equation}
Using
\begin{equation}
\frac{d}{dt}\left(\frac{1}{2}|{\bm{u}}|^2\right)=\bm{u}\cdot\dot{\bm{u}},
\qquad
\frac{d}{dt}(gz)=g\dot{z},
\end{equation}
we obtain
\begin{equation}
\frac{dE}{dt}
=
\frac{1}{m}\,\bm{u}\cdot\bm{F}_{\mathrm{aero}}
+
\frac{1}{m}\,\bm{u}\cdot\bm{F}_{\mathrm{muscle}}.
\end{equation}

We decompose the aerodynamic force into lift and drag,
\begin{equation}
\bm{F}_{\mathrm{aero}} = \bm{L} + \bm{D}.
\end{equation}
By definition,
\begin{equation}
\bm{L}\cdot\bm{v}=0,
\qquad
\bm{D}=-D\,\hat{\bm{v}},
\qquad
\hat{\bm{v}}=\frac{\bm{v}}{\abs{\bm{v}}},
\qquad
D\ge 0.
\end{equation}
Since $\bm{u}=\bm{v}+\bm{w}$, we have
\begin{equation}
\bm{u}\cdot\bm{L}=(\bm{v}+\bm{w})\cdot\bm{L}=\bm{w}\cdot\bm{L},
\end{equation}
and
\begin{equation}
\bm{u}\cdot\bm{D}
=
(\bm{v}+\bm{w})\cdot(-D\hat{\bm{v}})
=
-D\abs{\bm{v}}+\bm{w}\cdot\bm{D}.
\end{equation}
Therefore,
\begin{equation}
\bm{u}\cdot\bm{F}_{\mathrm{aero}}
=
\bm{w}\cdot(\bm{L}+\bm{D})-D\abs{\bm{v}}.
\end{equation}
Substituting into the energy balance yields
\begin{equation}
\frac{dE}{dt}
=
\frac{1}{m}\,\bm{w}\cdot(\bm{L}+\bm{D})
-
\frac{D\abs{\bm{v}}}{m}
+
\frac{1}{m}\,\bm{u}\cdot\bm{F}_{\mathrm{muscle}}.
\end{equation}
We therefore identify
\begin{equation}
\frac{dE}{dt}=P_{\mathrm{wind}}-P_{\mathrm{drag}}+P_{\mathrm{muscle}},
\end{equation}
with
\begin{equation}
P_{\mathrm{wind}}\equiv\frac{1}{m}\,\bm{w}\cdot(\bm{L}+\bm{D}),
\qquad
P_{\mathrm{drag}}\equiv\frac{D\abs{\bm{v}}}{m},
\qquad
P_{\mathrm{muscle}}\equiv\frac{1}{m}\,\bm{u}\cdot\bm{F}_{\mathrm{muscle}}.
\end{equation}
Dynamic soaring corresponds to the regime in which sustained transport is achieved primarily by positive environmental work $P_{\mathrm{wind}}$ rather than by large sustained muscular input.

\sisection{S2. Integrated work balance and empirical wind-harvest estimate}
Integrating the instantaneous energy balance over a time interval $[t_0,t_1]$ gives
\begin{equation}
E(t_1)-E(t_0)
=
\int_{t_0}^{t_1} P_{\mathrm{wind}}(t)\,dt
-
\int_{t_0}^{t_1} P_{\mathrm{drag}}(t)\,dt
+
\int_{t_0}^{t_1} P_{\mathrm{muscle}}(t)\,dt.
\end{equation}
Define
\begin{equation}
\Delta E \equiv E(t_1)-E(t_0),
\end{equation}
\chg{and we write the running relative energy as $\delta E(t) \equiv E(t)-E(t_0)$ (the quantity plotted in Fig.~1b).}
\begin{equation}
W_{\mathrm{wind}} \equiv \int_{t_0}^{t_1} P_{\mathrm{wind}}(t)\,dt,
\qquad
W_{\mathrm{drag}} \equiv \int_{t_0}^{t_1} P_{\mathrm{drag}}(t)\,dt,
\qquad
W_{\mathrm{muscle}} \equiv \int_{t_0}^{t_1} P_{\mathrm{muscle}}(t)\,dt.
\end{equation}
Then
\begin{equation}
\Delta E = W_{\mathrm{wind}} - W_{\mathrm{drag}} + W_{\mathrm{muscle}}.
\end{equation}
For windows dominated by gliding rather than sustained flapping, we approximate
\begin{equation}
W_{\mathrm{muscle}} \approx 0,
\end{equation}
which gives
\begin{equation}
W_{\mathrm{wind}} \approx \Delta E + W_{\mathrm{drag}}.
\end{equation}
\add{The quantity we can compute replaces the true drag by its ground-speed estimate, defining the harvest term $W_{\mathrm{harvest}} \equiv \Delta E + \widehat{W}_{\mathrm{drag}}$. This is the estimate of the true wind work $W_{\mathrm{wind}}$. The two coincide when muscular work is negligible and the drag model is exact, and $W_{\mathrm{harvest}}$ is otherwise an effective atmospheric input rather than a direct measurement of aerodynamic wind work.} $W_{\mathrm{harvest}}$ is the quantity plotted in Fig.~1c and used throughout the empirical energy balance. In the code, the signed drag contribution is stored as a negative cumulative quantity \chg{$W_{\mathrm{drag}}^{-}$},
\begin{equation}
\chg{W_{\mathrm{drag}}^{-}(t) = \int_{t_0}^{t} \left(-P_{\mathrm{drag}}(s)\right)\,ds,}
\end{equation}
so that the inferred harvest is implemented as
\begin{equation}
\chg{W_{\mathrm{harvest}}(t) = \delta E(t) - W_{\mathrm{drag}}^{-}(t).}
\end{equation}
Because \chg{$W_{\mathrm{drag}}^{-}(t)\le 0$}, this is algebraically equivalent to adding the magnitude of the dissipative loss.

\add{\textbf{Origin of the ledger quantities (Figs.~1b \& 1c).} The energy related to the ground frame is derived directly from the data, set by $E(t)=\tfrac{1}{2}u^2+gz$. This energy is calculated using the GPS Doppler ground speed $u$ and the recorded altitude $z$. As a result, the changes in energy, denoted as $\Delta E$ and $\delta E$, are quantities we can measure kinematically. 
\
On the other hand, the drag work, denoted as $W_{\mathrm{drag}}$, isn’t measured directly but is modeled. It comes from the integral of a quasi-steady gliding estimate, $\widehat{P}_{\mathrm{drag}}\approx g\,u/\mathcal{G}$ (S3), which combines the speed we've measured with a constant glide ratio $\mathcal{G}$ that we assume. 
\
As for the harvest term $W_{\mathrm{harvest}}$, it doesn’t have an independent measurement or model; instead, it's derived by closing the energy balance, assuming $W_{\mathrm{muscle}}\approx 0$. This gives us the relationship $W_{\mathrm{harvest}}=\Delta E+W_{\mathrm{drag}}$. So essentially, it is capturing the residual of atmospheric input rather than providing a direct measurement of the aerodynamic wind work. 
\
These quantities are primarily used in the energy-exchange panels shown in Fig.~1. However, for the cross-species comparison illustrated in Fig.~2, none of these quantities are used. Instead, the axes represent the normalized VeDBA effort $Y$ and the ground-speed ratio $X$ (S9--S10), with the coefficients $(a,b,W)$ derived from theoretical foundations and specifically fitted to the albatross frontier (S5).}

\sisection{S3. Drag model and species-specific parameters}
To estimate aerodynamic dissipation from trajectory data, we use a quasi-steady gliding approximation. \chg{Writing $L\equiv\abs{\bm{L}}$ and $D\equiv\abs{\bm{D}}$ for the lift and drag magnitudes, we define the lift-to-drag (glide) ratio}
\begin{equation}
\chg{\mathcal{G} \equiv \frac{L}{D} = \frac{\abs{\bm{L}}}{\abs{\bm{D}}},}
\end{equation}
so
\begin{equation}
\chg{D = \frac{L}{\mathcal{G}}.}
\end{equation}
For sustained gliding outside strong transients,
\begin{equation}
L \approx mg,
\end{equation}
hence
\begin{equation}
\chg{D \approx \frac{mg}{\mathcal{G}}.}
\end{equation}
The specific drag power is then
\begin{equation}
\chg{P_{\mathrm{drag}}
=
\frac{D\abs{\bm{v}}}{m}
\approx
\frac{g\abs{\bm{v}}}{\mathcal{G}}.}
\end{equation}
Because the trajectories are analyzed in the ground frame and the surrounding wind field is not reconstructed pointwise, we replace airspeed by the measured speed used in the ledger:
\begin{equation}
\chg{\widehat{P}_{\mathrm{drag}} \approx \frac{g\,u}{\mathcal{G}},
\qquad
u = \abs{\bm{u}}.}
\end{equation}
\chg{Throughout this SI, the hat symbol ($\widehat{\,\cdot\,}$) denotes a ground-frame estimate in which the unmeasured airspeed $\abs{\bm{v}}$ is replaced by the measured ground speed $u$.} \chg{Only the drag term carries this substitution directly. Quantities built from it (the harvest residual $W_{\mathrm{harvest}}$ and the window-averaged powers $\overline{P}_{\mathrm{harvest}}$ and $\overline{P}_{\mathrm{drag}}$) inherit the approximation, and we omit the hat on them for readability.}
In the implementation, this appears with a negative sign as
\begin{equation}
\chg{-\widehat{P}_{\mathrm{drag}} = -\frac{g\,u}{\mathcal{G}}.}
\end{equation}

\add{The species-specific value used in the energy-ledger drag model is $(L/D)_{\mathrm{alb}} = 21.2$, following Richardson et al.~\cite{richardson2018flight}. This value enters only the Fig.~1c energy ledger and the albatross-only comparison in SI\,S11, both applied to the wandering albatross alone. The cross-species comparison in Fig.~2 uses VeDBA (or the running 4-second wingbeat rate for the crane) as the effort axis and does not depend on species-specific $L/D$ values.}

\sisection{S4. Interpretation of ground speed versus airspeed}
The recorded kinematic quantities are fundamentally ground-referenced. \add{For the wandering albatross dataset, ground speed is the magnitude of the Doppler eastward and northward velocity components,}
\begin{equation}
u(t) = \sqrt{u_E^2(t)+u_N^2(t)},
\end{equation}
\add{and $z(t)$ is the recorded altitude. The empirical energy ledger is therefore a ground-frame energy balance,}
\begin{equation}
E(t)=\frac{1}{2}u^2(t)+gz(t).
\end{equation}
This does not eliminate the role of wind. Instead, it is precisely in the ground frame that a moving atmosphere can do nonzero work on the bird. What is not available in the present dataset-level analysis is a full reconstruction of the ambient wind field along each trajectory. For that reason, the quantity we infer from the ledger is best interpreted as an effective cumulative atmospheric input,
\begin{equation}
\chg{W_{\mathrm{harvest}} = \Delta E + \widehat{W}_{\mathrm{drag}},}
\end{equation}
rather than a direct pointwise measurement of the aerodynamic work of the wind field.

\sisection{S5. Derivation of the HJB limit}
We now derive the theoretical HJB limit curve in Figure 2a. Consider one-dimensional progress along a prescribed migration or return direction. Let $s(t)$ denote the remaining along-track distance to be covered. We take
\begin{equation}
\dot{s}(t) = -V(t),
\qquad V(t)>0,
\end{equation}
so the control variable is the transport speed $V(t)$. The quantity to be minimized is the total specific muscular work needed to traverse the remaining distance. Define the value function
\begin{equation}
J(s)
=
\inf_{V(\cdot)} \int_0^T p_{\mathrm{muscle}}\big(V(t)\big)\,dt,
\end{equation}
subject to
\begin{equation}
\dot{s}(t)=-V(t),
\qquad
s(0)=s,
\qquad
s(T)=0.
\end{equation}
Here $p_{\mathrm{muscle}}(V)$ is the specific muscular power required to sustain transport speed $V$ after accounting for aerodynamic losses and for energy harvested from the wind.

To obtain an analytically tractable reduced model, we decompose $p_{\mathrm{muscle}}$ into three parts.

First, induced drag decreases with increasing speed. Writing $D_i \propto 1/V^2$ in quasi-steady flight gives an induced-power contribution
\begin{equation}
p_i(V)=\frac{A}{V},
\end{equation}
where $A>0$ is an effective constant incorporating body weight and wing geometry.

Second, parasite drag increases quadratically with speed, $D_p \propto V^2$, so the corresponding power scales as
\begin{equation}
p_p(V)=B V^3,
\end{equation}
with $B>0$, an effective parasite-drag coefficient.

Third, the dynamic-soaring harvest term is taken to scale as
\begin{equation}
p_w(V)=\Gamma V^2,
\end{equation}
with $\Gamma>0$. This scaling follows from a simple cycle argument. If one dynamic-soaring cycle extracts an amount of energy proportional to the shear encountered over the cycle, then the energy gained per cycle scales as $\Delta E_{\mathrm{cyc}}\sim mV\Delta U$, where $\Delta U$ is an effective shear increment. If the spatial extent of a cycle is $\ell$, then the cycle frequency scales as $V/\ell$. The mean harvested power is therefore
\begin{equation}
P_{\mathrm{harvest}} \sim \Delta E_{\mathrm{cyc}}\times \frac{V}{\ell} \propto V^2.
\end{equation}
The coefficient $\Gamma$ absorbs the effective wind shear, cycle geometry, and conversion efficiency.

The net specific muscular power is therefore modeled as
\begin{equation}
p_{\mathrm{muscle}}(V)=\frac{A}{V}+BV^3-\Gamma V^2.
\end{equation}
Because the problem is stationary in the remaining distance variable $s$, the Hamilton--Jacobi--Bellman equation is
\begin{equation}
0 = \inf_{V>0}\left\{p_{\mathrm{muscle}}(V) + J_s(s)\,\dot{s}\right\}
= \inf_{V>0}\left\{p_{\mathrm{muscle}}(V)-VJ_s(s)\right\},
\end{equation}
with boundary condition
\begin{equation}
J(0)=0.
\end{equation}
In a homogeneous environment, the optimal cost is extensive in distance, so we set
\begin{equation}
J(s)=\lambda s,
\end{equation}
where $\lambda$ is the minimum specific energetic cost per unit distance. Then
\begin{equation}
J_s(s)=\lambda,
\end{equation}
and the HJB equation becomes
\begin{equation}
0 = \inf_{V>0}\left\{\frac{A}{V}+BV^3-\Gamma V^2-\lambda V\right\}.
\end{equation}
Dividing by $V>0$ gives
\begin{equation}
\lambda = \inf_{V>0}\left\{\frac{A}{V^2}+BV^2-\Gamma V\right\}.
\end{equation}
This expression is the reduced optimal-control bound. It states that the minimum cost per unit distance is obtained by minimizing the competition among an induced-drag term $A/V^2$, a parasite-drag term $BV^2$, and a wind-harvest term $-\Gamma V$. \chg{The division by $V$ converts specific power (units $\mathrm{m^2\,s^{-3}}$) into specific energetic cost per unit distance (units $\mathrm{m\,s^{-2}}$): each term on the right is now a cost per distance, and $\lambda$ is the minimum such cost. This is the quantity carried through to the plotted bound.}

The stationary point satisfies
\begin{equation}
\dv{}{V}\left(\frac{A}{V^2}+BV^2-\Gamma V\right)=0,
\end{equation}
that is,
\begin{equation}
-\frac{2A}{V^3}+2BV-\Gamma=0.
\end{equation}
Multiplying by $V^3$ yields
\begin{equation}
2BV^4-\Gamma V^3-2A=0,
\end{equation}
which determines the HJB-optimal transport speed in the reduced model.

To map this result onto the plotted reduced coordinates (the empirical reduced variables $X=V/V_{\mathrm{base}}$ and $Y$ are constructed from data in S10 below), we write
\begin{equation}
V = V_{\mathrm{base}} X
\end{equation}
and introduce a characteristic cost scale $C_0$. Then
\begin{equation}
\frac{1}{C_0}\left(\frac{A}{V^2}+BV^2-\Gamma V\right)
=
\frac{A}{C_0V_{\mathrm{base}}^2}\frac{1}{X^2}
+
\frac{BV_{\mathrm{base}}^2}{C_0}X^2
-
\frac{\Gamma V_{\mathrm{base}}}{C_0}X.
\end{equation}
Defining
\begin{equation}
a \equiv \frac{A}{C_0V_{\mathrm{base}}^2},
\qquad
b \equiv \frac{BV_{\mathrm{base}}^2}{C_0},
\qquad
W \equiv \frac{\Gamma V_{\mathrm{base}}}{C_0},
\end{equation}
we obtain the \chg{(unclipped)} reduced HJB form
\begin{equation}
\chg{\widetilde{Y}_{\mathrm{HJB}}(X)=\frac{a}{X^2}+bX^2-WX,}
\end{equation}
\chg{where $C_0$ is an arbitrary characteristic cost scale that is absorbed into the fitted coefficients $(a,b,W)$ and is never determined independently. Both the theoretical bound and the empirical ordinate $Y=(\mathcal{E}-\mathcal{E}_0)/\mathcal{E}_0$ are dimensionless: the former through division by the characteristic cost scale $C_0$, the latter through division by the effort baseline $\mathcal{E}_0$. The comparison in Fig.~2 is therefore between two dimensionless quantities sharing a common reduced speed axis $X$, not between a physical energy and a physical power. The theory enters through the functional form of the bound, not through absolute units.}
The displayed curve is clipped at zero,
\begin{equation}
\chg{Y_{\mathrm{HJB}}(X)=\left[\widetilde{Y}_{\mathrm{HJB}}(X)\right]_{+}=\max\!\left(0,\;\frac{a}{X^2}+bX^2-WX\right),}
\end{equation}
\chg{which reflects that $Y_{\mathrm{HJB}}$ is a normalized excess-effort quantity relative to a low-effort baseline.} Negative reduced costs are therefore not displayed as negative effort. \add{The parameters $(a,b,W)$ are estimated on the albatross frontier by constrained sequential quadratic programming (SLSQP), minimizing the least-squares residual subject to $Y_{\mathrm{HJB}}(x_i) \le Y_{\mathrm{frontier}}(x_i)$ at every frontier point and to $a,b \ge 10^{-6}$ and $W \ge 0.05$ (the physical floor; see main text). The fitted values are $a = 0.0101 \pm 0.0020$, $b = 0.1127 \pm 0.0040$, and $W = 0.0577 \pm 0.0056$. $W$ reaches the floor in nine of the eleven cross-validation folds below, so the constraint is active on subsets of the data and nearly inactive on the full fit. The calibration is validated by 11-fold cross-validation on the albatross deployments: out-of-sample median $\Delta Y = +0.152$ (IQR $[+0.066,\,+0.250]$) versus in-sample median $+0.132$ (IQR $[+0.049,\,+0.213]$), a mean absolute difference of $0.021$ per deployment (\texttt{kfold\_albatross.py}).} Accordingly, our reduced HJB-limit is justified by the form of the reduced optimal-control derivation above, while the numerical parameter values are determined empirically from the data.

\sisection{S6. Datasets}
\paragraph{Wandering albatrosses.}

\add{Dryad archive ``Behavioral datasets of wandering albatrosses collected at Possession Island, Crozet, France, in 2019 and 2020'' (\texttt{10.5061/dryad.tx95x6b2j}). Ninja-scan recorders operated at $100\,$Hz (ACC) and $5\,$Hz (GPS). In 2019, 21 individuals were tracked, and in 2020, 23, resulting in 44 valid individuals across 55 deployments (11 individuals provided two foraging trips each). Each deployment was normalized separately.}

\paragraph{Short-tailed shearwater.}

\add{Data for \textit{Ardenna tenuirostris} was contributed by John Arnould from Griffith Island and Gabo Island, Australia~\cite{berlincourt2015combined} and was provided via the Zenodo archive (\texttt{10.5281/zenodo.21304763}). Foraging occurred under natural circumstances without experimental manipulations. From continuous $25\,$Hz triaxial accelerometry, coupled with 5-minute GPS fixes for 10 individuals providing 13 foraging trip deployments, there were 12,811 windows in this data set. Ground speed was obtained from the logger's GPS speed channel, converted from km/h, and then linearly interpolated to the accelerometry data timeline. This results in logged per-fix speeds rather than haversine-estimated distances divided by time between fixes. Since this species rests on the water for approximately $46\%$ of its time at sea~\cite{berlincourt2015combined}, a speed threshold was applied to exclude windows spent resting or drifting. An $8\,$m/s speed gate was applied yielding a calculated $V_{\mathrm{base}} = 11.19\,$m/s.}

\paragraph{Buller's albatross.}

\add{\textit{Thalassarche bulleri bulleri} (southern Buller's albatross) data were collected from Hautere/Solander Island, New Zealand, during the 2016 and 2017 chick-rearing periods~\cite{poupart2019bullers}. The GPS tracks, archived in the BirdLife Seabird Tracking Database (dataset 2081), are paired with $25\,$Hz accelerometry data contributed by J. Arnould and stored at Zenodo (\texttt{10.5281/zenodo.21304763}). Continuous $25\,$Hz accelerometry, combined with GPS at a 2-minute fix interval~\cite{waugh2017muttonbirders}, yielded 32 deployments and 132,321 windows. Ground speed is calculated by finding the difference in haversine distances of consecutive GPS fixes (there is no explicit speed channel in the raw export). Since the bird does not fly in a straight line between fixes taken two minutes apart, this can lead to an underestimation of true ground speed. The median ground speed in high-effort windows is $8.46\,$m/s, compared to published mollymawk cruising speeds of $13$--$15\,$m/s, leading to a compression of about 1.6-fold. This compression closely represents a multiplicative factor on $V$, and $V_{\mathrm{base}}$ is the median of the same compressed record. Therefore, it largely cancels in $X = V/V_{\mathrm{base}}$. The measured value reported is $V_{\mathrm{base}} = 9.51\,$m/s.}

\paragraph{Common crane.}

\add{\textit{Grus grus} The GPS and accelerometry data comes from Pekarsky et al.~\cite{pekarsky2024cranes} (Dryad \texttt{10.5061/dryad.t76hdr871}) and includes 32 individuals at $1\,$Hz. Since $1\,$Hz cannot capture individual wingbeats, the running 4-second wingbeat rate (\texttt{running\_Flap\_rate\_4sec}) was used as the effort proxy instead of VeDBA. Because $Y = (\mathcal{E}-\mathcal{E}_0)/\mathcal{E}_0$ is a within-species ratio, the difference in effort units does not impact comparisons.}

\sisection{S7. Preprocessing for the albatross population-level analysis}
\add{For the wandering albatross analysis, the acceleration and GPS streams are aligned on the GPS time base, and GPS fixes with zero latitude or longitude are discarded.} The median positive GPS time increment defines the effective GPS sampling interval
\begin{equation}
\Delta t_{\mathrm{GPS}} = \mathrm{median}\left\{t_{k+1}-t_k : t_{k+1}>t_k\right\}.
\end{equation}
Acceleration components are then linearly interpolated onto the GPS times.

Dynamic body acceleration is computed by subtracting a low-frequency gravitational component from each interpolated acceleration channel. Specifically, each channel is low-pass filtered with a fourth-order Butterworth filter at the cutoff frequency
\begin{equation}
f_c = 0.2\ \mathrm{Hz},
\end{equation}
and the dynamic component is defined as the residual after subtraction. The resulting vectorial dynamic body acceleration is
\begin{equation}
\mathrm{VeDBA}(t)
=
\sqrt{\left(a_x-a_x^{\mathrm{lp}}\right)^2
+
\left(a_y-a_y^{\mathrm{lp}}\right)^2
+
\left(a_z-a_z^{\mathrm{lp}}\right)^2 }.
\end{equation}

Ground speed is reconstructed from the Doppler eastward and northward components,
\begin{equation}
u(t)=\sqrt{u_E^2(t)+u_N^2(t)},
\end{equation}
and heading is
\begin{equation}
\psi(t)=\mathrm{atan2}\!\left(u_N(t),u_E(t)\right).
\end{equation}
The turn rate is then computed from the unwrapped heading by finite differentiation,
\begin{equation}
\dot{\psi}(t) \approx \frac{d}{dt}\,\mathrm{unwrap}\,\psi(t).
\end{equation}

\sisection{S8. Windowing and window-level observables}
Continuous albatross trajectories are partitioned into overlapping windows of duration
\begin{equation}
T_w = 120\ \mathrm{s}
\end{equation}
with step size
\begin{equation}
\Delta T = 30\ \mathrm{s}.
\end{equation}
\add{A 120-s window was used to enclose several full dynamic-soaring cycles, typically between 7 and 30 s ($0.03$--$0.15\,$Hz), but short enough to likely contain a single persistent behavioral state. To show that this decision did not create the observed outcome, we reran the analysis using a 60-s window for the wandering albatross and a 240-s window, keeping the comparison species at 120 s. In all three window lengths, species order remains the same (wandering albatross, Buller's albatross, short-tailed shearwater, common crane); comparison-species median $\Delta Y$ varies $\le$6\% across window lengths (albatross $\approx$20\%, reflecting closeness to the bound); and separation varies between 22.6--28.3$\times$ between the wandering albatross and common crane. Re-estimation of the bound at each window length for each re-run means absolute values differ, but results are robust (e.g., here, wandering albatross median $\Delta Y=0.176$ vs main 0.132 (Fig.~2b) behind a 33$\times$ separation here and in the main albatross-crane range, between $\approx 23\times$--$33\times$ separation).}

For the albatross, a window is retained only if at least $90\%$ of the speed values in that window are finite and if the mean ground speed satisfies
\begin{equation}
\overline{u} \ge 6.0\ \mathrm{m/s}.
\end{equation}
For energy-ledger quantities, \add{we additionally require} that at least $80\%$ of altitude values within the window be finite.

\add{For each retained window we compute} a transport-progress statistic
\begin{equation}
\mathrm{progress}
=
\frac{1}{T_w}
\sqrt{
\left(\sum_{k\in w} u_E(t_k)\,\Delta t\right)^2
+
\left(\sum_{k\in w} u_N(t_k)\,\Delta t\right)^2
},
\end{equation}
the mean dynamic body acceleration
\begin{equation}
\overline{\mathrm{VeDBA}}
=
\frac{1}{N_w}\sum_{k\in w}\mathrm{VeDBA}(t_k),
\end{equation}
and a dynamic-soaring spectral intensity defined as the Welch bandpower of the turn-rate signal in the frequency band
\begin{equation}
f \in [0.03,\,0.15]\ \mathrm{Hz}.
\end{equation}
That is,
\begin{equation}
P_{\mathrm{DS}}
=
\int_{0.03}^{0.15} S_{\dot{\psi}}(f)\,df,
\end{equation}
where $S_{\dot{\psi}}(f)$ is the Welch estimate of the turn-rate power spectral density.

Within each window, the energy ledger uses a Savitzky--Golay filter with polynomial order
\begin{equation}
p=2
\end{equation}
and window length
\begin{equation}
T_{\mathrm{SG}} = 3.0\ \mathrm{s}
\end{equation}
applied separately to speed and altitude before constructing
\begin{equation}
E(t)=\frac{1}{2}u_{\mathrm{sm}}^2(t)+gz_{\mathrm{sm}}(t).
\end{equation}
\add{Three window-level energy quantities are recorded:}
\begin{equation}
\overline{P}_{\mathrm{harvest}}
=
\frac{W_{\mathrm{harvest}}(t_1)-W_{\mathrm{harvest}}(t_0)}{T_w},
\end{equation}
\begin{equation}
\overline{P}_{\mathrm{drag}}
=
\frac{W_{\mathrm{drag}}(t_1)-W_{\mathrm{drag}}(t_0)}{T_w},
\end{equation}
and
\begin{equation}
\chg{\Delta E
=
\delta E(t_1)-\delta E(t_0).}
\end{equation}
\add{Both $\overline{P}_{\mathrm{harvest}}$ and $\overline{P}_{\mathrm{drag}}$ are reported as non-negative rates, each the increment of a positive cumulative quantity ($W_{\mathrm{harvest}}$ and $\widehat{W}_{\mathrm{drag}}$, respectively) over the window duration $T_w$. The signed convention $W_{\mathrm{drag}}^{-}\le 0$ (S2) enters only the intermediate harvest sum and does not propagate to these rates. $\Delta E$ is signed by definition.}

\sisection{\chg{S8.1 Windowing by species}}

\chg{All four species use the same $120$-second analysis window stepped by $30$ seconds (Eqs.~62--63). The species differ only in raw sensor rate and in how the window-level speed and effort are formed.}

\paragraph{\chg{Wandering albatross.}}

\chg{The Ninja-scan samples at 100 Hz (accelerometry) and 5 Hz (Doppler GPS). Each window of 120 seconds duration will include 600 GPS samples. Speed is the mean Doppler ground speed, and effort is the mean VeDBA. A window is kept if at least 90\% of the speed measurements are finite and the mean ground speed over the window is at least 6 m/s (Eq.~64). Similar conditions must hold for altitude values to keep energy-ledger measurements finite (at least 80\%).}

\paragraph{\chg{Buller's albatross.}}

\chg{Accelerometry is provided at 25 Hz. The GPS information is provided with a 2-minute resolution, meaning each window will consist of about one single raw GPS reading. Ground speed is calculated by the haversine difference between consecutive readings, and this reading is used in each of the subsequent 2 minutes within the window. This 2-minute fix spacing reduces the average measured speed by a factor of about 1.6 relative to the median speed of the flight (S6). Effort is the average VeDBA over the window.}

\paragraph{\chg{Short-tailed shearwater.}}

\chg{Accelerometry is at 25 Hz. GPS fixes are taken every 5 minutes. Ground speed is obtained from the logger's recorded GPS speed channel (converted from km/h) and is interpolated onto the accelerometer timeline while the effort is the average VeDBA. This species spends about 46\% of its time at sea resting on the water, so an 8 m/s flight gate is used when calculating $V_{\rm base}$ (S6).}

\paragraph{\chg{Common crane.}}

\chg{GPS and flap rate are recorded once per second. So, every 120-second window has 120 GPS and flap-rate samples. Ground speed represents the mean ground speed during that time, and effort is the mean flap rate from S10.1 (4s running avg) during that time. Windows where the mean ground speed is less than 3 m/s are discarded.}

\chg{In each instance, the speed and effort at the window level are condensed into a single value per window, resulting in the reduced variables $X$ and $Y$ being constructed in the same way, regardless of the initial sampling resolution across species.}

\sisection{S9. Empirical Pareto construction for the effort--progress plane}
The population-level Pareto plane used for the albatross analysis (Figure 1d) takes
\begin{equation}
X_{\mathrm{emp}} = \mathrm{progress}_{\mathrm{m/s}},
\qquad
Y_{\mathrm{emp}} = \overline{\mathrm{VeDBA}}.
\end{equation}
\add{After dropping non-finite observations, the lower envelope is estimated by the 10th conditional quantile of $Y{\mathrm{emp}}$ on $X{\mathrm{emp}}$ using quantile regression on a cubic B-spline basis (quantile $\tau=0.10$, an estimator of the same class as those that produce the cross-species frontiers in Figure 2 (using $\tau=0.25$). A lower quantile is used because Figure 1d uses units that are not normalized per species, so the lack of a per-species minimum tends to force the frontier downwards to $Y<0$. There are six interior knots. We make no use of this fitted frontier beyond presenting a smooth rendering of the empirical lower envelope.}

\sisection{S10. Comparative reduced phase space}
\add{A secondary representation compares the four species in a shared reduced phase space (Fig.~2a). \chg{For the procellariiform comparison species (Buller's albatross, short-tailed shearwater), data are derived from GPS-accelerometer-merged $25\,$Hz files.} VeDBA is calculated according to methods outlined in S8: the gravity-static component is removed via the same fourth-order $0.2\,$Hz Butterworth low-pass filter and taken as the vector magnitude of the remainder. Buller's albatross speeds, computed from haversine differences between 2-minute GPS locations, were additionally smoothed with a $5\,$s moving average (125 samples @ $25\,$Hz). The short-tailed shearwater uses speeds directly from the logger's built-in GPS speed record (converted from km/h). This smoothing has no effect on anything other than speeds. The common crane is compared using a measure of effort; specifically, the running 4-second wingbeat rate \texttt{running\_Flap\_rate\_4sec} sampled at $1\,$Hz.}

After the merge, rows are retained only if
\begin{equation}
\add{1.0 < V_{\mathrm{obs}} < 30.0\ \mathrm{m/s}.}
\end{equation}
\add{For the short-tailed shearwater that spent ${\sim}46\%$ of its at-sea time on the surface of the water, a further minimal speed of $8\,$m/s has been used when calculating $V_{\mathrm{base}}$, which excludes windows where it was drifting from the cruise-speed reference (see SI\,S6). Since $Y$ is an at-sea relative effort ratio at the within-species level, the species-specific conversion from effort to power will cancel out in the numerator and denominator, as it is a multiplicative factor.}
For a given species dataset, let $V_{\mathrm{obs}}$ denote the retained ground-speed values and let $\mathcal{E}_{\mathrm{obs}}$ denote the corresponding VeDBA values. The reduced variables are then defined by
\begin{equation}
\mathcal{E}_0 = \mathrm{percentile}_{5}(\mathcal{E}_{\mathrm{obs}}),
\end{equation}
\begin{equation}
\chg{V_{\mathrm{base}} = \mathrm{median}\left\{V_{\mathrm{obs}} : \mathcal{E}_{\mathrm{obs}} < \mathrm{percentile}_{15}(\mathcal{E}_{\mathrm{obs}})\right\}.}
\end{equation}
\begin{equation}
X = \frac{V_{\mathrm{obs}}}{V_{\mathrm{base}}},
\qquad
Y = \frac{\mathcal{E}_{\mathrm{obs}}-\mathcal{E}_0}{\mathcal{E}_0}.
\end{equation}
Thus, $X$ is a reduced transport speed and $Y$ is a normalized excess effort above a low-effort baseline.

For the albatross population, \add{the same transformation is applied per deployment to the window-level transport-progress and mean-VeDBA values.} We first restrict to values satisfying
\begin{equation}
1 < \mathrm{progress} < 35,
\qquad
0.1 < \overline{\mathrm{VeDBA}} < 10,
\end{equation}
and then sets
\begin{align}
\mathcal{E}_0 &= \mathrm{percentile}_{5}(\overline{\mathrm{VeDBA}}), \\
V_{\mathrm{base}} &= \mathrm{clip}\!\left(
\mathrm{median}\left\{\mathrm{progress} : \overline{\mathrm{VeDBA}} < \mathrm{percentile}_{15}(\overline{\mathrm{VeDBA}})\right\},\,10,\,20\right).
\end{align}

\add{The frontier for each species is the 25th percentile of $Y$ given $X$, estimated via quantile regression with a cubic B-spline basis ($\tau=0.25$). The 25th percentile was chosen because, after normalization using each species’ own baseline, the least-costly windows expand to form a band that dips below zero. The 10th percentile could dip below zero for high $X$, whilst $Y_{\mathrm{HJB}}$ must be positive. Six interior knots are used. The frontiers are displayed over the $X$-range, within which at least one window of each species lies within the support for which albatross data exist without any extrapolation.}

\sisection{\chg{S10.1 Effort proxy for the common crane}}

\chg{At $1\,$Hz the crane accelerometry cannot resolve wingbeats, so VeDBA is unavailable. In its place $\mathcal{E}$ is the \texttt{running\_Flap\_rate\_4sec} channel (flaps/s) supplied with the Pekarsky et al.~\cite{pekarsky2024cranes} dataset, a $4$-second running wingbeat rate calibrated against $10\,$Hz accelerometry. Processing otherwise follows the seabirds. We take $120\,$s windows stepped by $30\,$s, let $\mathcal{E}$ and $V$ be the window means, discard windows below $3\,$m/s, and recover $\mathcal{E}_0$ and $V_{\mathrm{base}}$ from Eqs.~(77)--(78). Averaging over $120\,$s smooths the flap count, the $3\,$m/s gate removes perched birds, and taking the $5$th percentile rather than the minimum keeps the baseline off the floor. What we measure is $\mathcal{E}_0 = 0.32$ flaps/s, with $V_{\mathrm{base}} = 14.1\,$m/s coming from these low-flap windows, which are cranes gliding between thermals rather than sitting still. And since $Y$ is a within-species ratio, moving from VeDBA to flaps/s cancels out and leaves the comparison unchanged.}

\sisection{S11. Construction of the albatross energy-balance panels (Fig.~1b,c)}
\add{The energy-balance panels of Fig.~1b,c are constructed for the wandering albatross alone. For each window, ground speed $u$ (the magnitude of the Doppler velocity components) and altitude $z$ are read from the record.} Rows with implausible values are removed using the hard thresholds
\begin{equation}
0 \le u < 50\ \mathrm{m/s},
\qquad
-50 < z < 200\ \mathrm{m}.
\end{equation}
If fewer than 60 seconds of data remain at the nominal sampling rate, the window is discarded. Otherwise, only the first 120 seconds are retained. Speed and altitude are then smoothed with a Savitzky-Golay filter of order 2 and a window length of approximately 5 seconds before computing
\begin{equation}
E(t)=\frac{1}{2}u_{\mathrm{sm}}^2(t)+gz_{\mathrm{sm}}(t),
\end{equation}
\begin{equation}
\chg{W_{\mathrm{drag}}^{-}(t)=\sum_{k\le t}\left(-\frac{g\,u_{\mathrm{sm}}(t_k)}{\mathcal{G}}\right)\Delta t,}
\end{equation}
\begin{equation}
\chg{W_{\mathrm{harvest}}(t)=\delta E(t)-W_{\mathrm{drag}}^{-}(t).}
\end{equation}
The mean harvested power for each window is
\begin{equation}
\overline{P}_{\mathrm{harvest}} = \frac{W_{\mathrm{harvest}}(t_{\mathrm{end}})}{t_{\mathrm{end}}}.
\end{equation}

\sisection{S12. Limitations and scope of interpretation}
\add{The three main limitations of the framework are mentioned in the Concluding Remarks of the main text. Additionally, we note that the raw-sensor pipeline for the wandering albatross (Ninja-scan GPS + ACC at 5/100 Hz) is different from those used for the comparison datasets. Buller's albatross and the short-tailed shearwater use 25 Hz accelerometry with GPS at 2- and 5-minute fix intervals. This means the Buller's speed axis is calculated using haversine, while the shearwater relies on the logger's recorded GPS speed channel. The crane uses 1 Hz GPS + ACC with wingbeat rate instead of VeDBA, but all datasets provide comparable reduced variables $(X, Y)$. However, differences in sensor type, sampling rate, and GPS fix interval still exist (see SI S6).}



\printbibliography